\documentclass[a4paper,12pt]{article}
\usepackage{amsfonts}
\usepackage{amssymb}
\usepackage{amsmath}
\usepackage{indentfirst}
\usepackage{dsfont}
\usepackage[latin1]{inputenc}
\def\be{\begin{equation}}
\def\ee{\end{equation}}
\def\bea{\begin{eqnarray}}
\def\eea{\end{eqnarray}}

\title{Dirac Equation in $\kappa$-Minkowski space-time}
\author{Ravikant Verma\footnote{Email:ravikant.uohyd@gmail.com}~\\
{\it School of Physics, University of Hyderabad},\\{\it Central University P O, Hyderabad-500046, India}}

\begin{document}
\maketitle
\begin{abstract}
In this paper, we derive the Dirac equation in the $\kappa$-deformed Minkowski space-time. We start with $\kappa$-deformed Minkowski space-time and investigate the undeformed $\kappa$-Lorentz transformation valid to all order in the deformation parameter $a$. Using the undeformed $\kappa$-Lorentz algebra, we obtain the $\kappa$-deformed Dirac equation, valid to all order in the deformation parameter $a$. In limit $a\rightarrow$0, we get back the correct commutative result.
\end{abstract}
\section{Introduction}

The noncommutative geometry has attracted wide attention in the context of study of space-time at the microscopic level\cite{1,2,3,4,5,6,7,8}. In the noncommutative geometry, the structure of the space-time is modified in comparison with the commutative space-time. The $\kappa$-Minkowski space-time is one of the examples of the space-time whose coordinates satisfy Lie algebra type commutation relations. Here, the deformed commutation relations between the space-time coordinates depend on the deformation parameter $a$ which has the dimension of length. This deformation of the space-time structure also necessitate the modification of the associated symmetry algebra. The modified symmetry algebra corresponding to the $\kappa$-Minkowski space-time is known as deformed-$\kappa$-Poincare algebra. $\kappa$-Minkowski space-time is also connected to the doubly special relativity(DSR)\cite{gl}, which introduces a fundamental constant of length dimension in addition to the constant velocity of light present in the special theory of relativity.

In \cite{19,20}, a $\kappa$-deformed Dirac equation compatiable with doubly special relativity was constructed in momentum space with specific choice of differential calculus defined on $\kappa$-Minkowski space-time. Another, $\kappa$-Dirac equation was constructed in \cite{21} by demanding that the square of this should be the $\kappa$-deformed Klein-Gordon equation. The $\kappa$-Dirac equation obtained in\cite{21} is invariant under the parity and time reversal operations but it was not invariant under the charge conjugation. By using this $\kappa$-Dirac equation, the modification of the Unruh effect was studied in \cite{22} (and for scalar field this was studied in \cite{23}). The Dirac equation in $\kappa$-deformed space-time was constructed and analysed by using different approaches in \cite{24,25,26,27,28}.

In this paper, we derive the $\kappa$-deformed Dirac equation valid to all orders in the deformation parameter, using a different approach. Starting from the deformed dispersion relation valid in the $\kappa$-deformed space-time, we first obtain the modified boost factor $\gamma^\prime$ exactly, i.e., valid to all orders in the deformation parameter. Using this $\gamma^\prime$ and the explicit form of the boost generators of undeformed $\kappa$-Poincare algebra, we then obtain finite boost transformations. Using the corresponding transformation matrix, we then derive the matrix realisation of boost generators. All these explicit matrix realisations are ${\it independent}$ of the deformation parameter. The deformation parameter appear only through the boost factor $\gamma^\prime$ and $\gamma^\prime$ is valid to all orders in the deformation parameter. Note that the rotation sector of the undeformed Poincare algebra are unaffected by the $\kappa$-deformation. This should not be surprising as the $\kappa$-deformation leave the spatial coordinates to be mutually commuting with each other.

As in the commutative case, by a change of basis, we re-express the undeformed $\kappa$-Lorentz algebra as the direct product of $su(2)$ algebras. Following the usual approach in the commutative space-time\cite{ryder}, we then define left and right $su(2)$ spinors. Using the explicit form of undeformed $\kappa$-Lorentz transformations, we obtain the relation between these $su(2)$ spinors with zero momentum and non-zero momentum $\vec{P}$. From this transformation rule, as in the commutative space-time, one gets the relation between the left and right spinors with arbitrary three momentum and energy. This relation written in the matrix form is nothing but the $\kappa$-deformed Dirac equation, in the momentum space. With the operator realisation of four momentum used in writing down the energy momentum relation, we show that the $\kappa$-deformed Dirac equation we obtained is mapped to the $\kappa$-Dirac equation obtained in\cite{21}, in the coordinate space. Further, we show that square of this Dirac equation correctly give the Klein-Gordon equation.

This paper is organized as follows. In next section we give a brief summary of the $\kappa$-Minkowski space-time and in section 3, we derive the Lorentz transformation in $\kappa$-Minkowski space-time which is required for deriving the Dirac equation in $\kappa$-Minkowski space-time. In section 4, we derive the Dirac equation in $\kappa$-Minkowski space-time using the $\kappa$-deformed Lorentz algebra. We give our concluding remarks in section 5.

\section{$\kappa$-Minkowski space-time}
In this section, first we recall the essential detail of the $\kappa$-deformed space-time, whose coordinates obey commutation relations,
\begin{equation}
 [\hat{x}^i, \hat{x}^j]=0,~~ [\hat{x}^0, \hat{x}^i]=i a \hat{x}^i,~~  a=\frac{1}{\kappa}.\label{kappacom}
\end{equation}
Here, $a$ is the deformation parameter and has the dimension of length. In the Minkowski space-time with metric $\eta_{{\mu}{\nu}}$ = diag(-1,1,1,1) we define $x^\mu =\eta^{{\mu}{\alpha}} x_\alpha$ and $\partial^{\mu}=\eta^{{\mu}{\alpha}} \partial_\alpha$. These coordinates and derivatives satisfy the commutation relations
\begin{eqnarray}
&[x_\mu, x_\nu]=0 ,~~ [\partial_\mu, \partial_\nu]=0 ,~~ [\partial^{\mu} , x_{\nu}]=\eta^{\mu} _{\nu},&\\
&[\partial_\mu, x_\nu]=\eta_{{\mu}{\nu}}, [p_\mu , x_\nu]=- i \eta_{{\mu}{\nu}},p_\mu=- i \partial_\mu .&
\end{eqnarray}
The $\kappa$-Minkowski space-time coordinates can be defined in terms the commutative coordinates $x_\mu$ and corresponding derivatives $\partial_\mu$ as $\hat{x}_\mu=x^\alpha \phi_{{\alpha}{\mu}}(\partial)$.
 In this realization, the explicit form of $\hat{x}_i$ and $\hat{x}_0$ are given as
\begin{eqnarray}
\hat{x}_i=x_i \varphi(A),~~ \hat{x}_0=x_0 \psi(A) + i a x_i \partial_i \gamma(A),\label{partial}
\end{eqnarray}
where $A=- ia \partial_0$. Using Eqn.(\ref{partial}) in Eqn.(\ref{kappacom}), we obtain
\begin{equation}
\frac{\varphi^\prime}{\varphi}\psi=\gamma(A)-1
\end{equation}
where $\varphi^\prime=\frac{d\varphi}{dA}$ satisfying the boundary conditions $\varphi(0)=1, \psi(0)=1, \gamma(0)=\varphi^\prime (0)+1$ and is finite and $\varphi, \psi, \gamma$ are positive functions.

 These coordinate satisfy the condition that in particular $[\partial_\mu , \hat{x}_\nu]=\phi_{{\mu}{\nu}}(\partial)$, we find 
\begin{equation}
[\partial_i , \hat{x}_j]=\delta_{{i}{j}} \varphi(A) ,~~[\partial_i , \hat{x}_0]=i a \partial_{i} \gamma(A) ,~~ [\partial_0 , \hat{x}_i]=0,
\end{equation}
and
\begin{equation}
[\partial_0 ,\hat{x}_0]=-\psi(A).
\end{equation}
Let $M_{\mu\nu}$ denote the rotation and boost generators of the $\kappa$-Poincare algebra. We require that the commutation relations between $M_{\mu\nu}$ should be same as the usual Poincare algebra and also that between $M_{\mu\nu}$ and the $\kappa$-space-time coordinates $\hat{x}_{\mu}$ should be linear functions of $\hat{x}_{\mu}$,~$M_{\mu\nu}$ and deformation parameter $a_\mu$. We also demand that these generators have the correct commutative limit. Thus one get
\begin{equation}
[M_{\mu\nu},\hat{x}_{\lambda}]=\hat{x}_{\mu}\eta_{\nu\lambda}-\hat{x}_{\nu}\eta_{\mu\lambda}-i(a_\mu M_{\nu\lambda}-a_\nu M_{\mu\lambda}).
\end{equation}
We also demand that these commutators have the correct commutative limit. This leads to two type of possible realisation, one where $\psi(A)=1$ and second one $\psi(A)=1+2A$\cite{fgrd1}. The generators of the undeformed $\kappa$-Poincare algebra are $D_\mu$ and $M_{\mu\nu}$, whose explicit forms (for the realization $\psi=1$) are given by
\begin{equation}
M_{ij}=x_{i}\partial_{j}-x_{j}\partial_{i},\label{dg1}
\end{equation}
\begin{equation}
M_{i0}=x_i \partial_0 \varphi \frac{e^{2A}-1}{2A}-x_0 \partial_i \frac{1}{\varphi}+iax_i \bigtriangledown^2 \frac{1}{2\varphi}-iax_k \partial_k \partial_i \frac{\gamma}{\varphi},\label{dg2}
\end{equation}
and
\begin{equation}
D_i=\partial_i \frac{e^{-A}}{\varphi},~~ D_0=\partial_0 \frac{sinhA}{A} + i a \bigtriangledown^2 \frac{e^{-A}}{2 \varphi^2}. \label{dr}
\end{equation}
Where  $\bigtriangledown^2=\partial_k \partial_k$.
The symmetry algebra of the underlying $\kappa$-space-time generated by $M_{\mu \nu}$ and $D_\mu$ is given above known as the undeformed $\kappa$-Poincare algebra. Their generators $D_{\mu}$ and $M_{{\mu}{\nu}}$ obey\cite{fgrd1,fgrd2,fgrd3,fgrd4,fgrd5}
\begin{equation}
[M_{{\mu}{\nu}} , D_{\lambda}]=\eta_{{\nu}{\lambda}} D_\mu - \eta_{{\mu}{\lambda}} D_\nu ,~~ [D_\mu , D_\nu]=0,\label{dirac4}
\end{equation}
\begin{equation}
[M_{{\mu}{\nu}}, M_{{\lambda}{\rho}}]=\eta_{{\mu}{\rho}} M_{{\nu}{\lambda}} + \eta_{{\nu}{\lambda}} M_{{\mu}{\rho}} - \eta_{{\nu}{\rho}} M_{{\mu}{\lambda}} - \eta_{{\mu}{\lambda}} M_{{\nu}{\rho}}.\label{dirac5}
\end{equation}
Here we note that the usual derivative $\partial_\mu$ do not transform like a 4-vector under the above defined undeformed $\kappa$-Poincare algebra whereas the Dirac derivative $D_\mu$ transform like a 4-vector. 
 
The undeformed $\kappa$-Poincare algebra in Eqns.(\ref{dirac4},\ref{dirac5}) have same form as the Poincare algebra. But the explicit form of generators are modified compared to that of the Poincare algebra. The explicit form of these generators are given in Eqn.(\ref{dg1}), Eqn.(\ref{dg2}) and Eqn.(\ref{dr}).

The Casimir of undeformed $\kappa$-Poincare algebra is $D_\mu D^\mu$ which in the momentum space is given by $P_\mu P^\mu$ and the dispersion relation in $\kappa$-Minkowski space-time is given as
\be 
P_\mu P^\mu =P_0 P^0 + P_i P^i = -m_0^2 .\label{1}
\ee
Note that the momentum $P_0$ and $P_i$ appearing in the above are $\kappa$-deformed momenta and they are different from the momenta in the commutative space-time. Explicitly, the above deformed dispersion relation is given as
\begin{equation}
\frac{4}{a^2}\sinh^2\left(\frac{ap_0}{2}\right)-p_{i}^2\frac{e^{-ap_0}}{\varphi^2(ap_0)} -\frac{a^2}{4}\left[\frac{4}{a^2}\sinh^2\left(\frac{ap_0}{2}\right)-p_{i}^2\frac{e^{-ap_0}}{\varphi^2(ap_0)}\right]^2=m_0^2,\label{dis}
\end{equation}
where $p_0$ and $p_i$ are the momenta in the commutative space-time.
From Eqn.(\ref{1}), we get
\be 
P_0^2 = E^{\prime 2} = |\vec{P}|^2 +m_0^2,
\ee
where $m_0$ is the rest mass of the particle. Here we note that $E^\prime$ appearing above is the deformed total energy of the particle in the $\kappa$-Minkowski space-time. It is given by $E^\prime = \sqrt{|\vec{P}|^2 + m_0^2}$, here $\vec{P}$ is the deformed three momentum of particle in the $\kappa$-Minkowski space-time. As in the commutative space-time, we write the boost factor $\gamma^\prime$ in $\kappa$-Minkowski space-time as $\gamma^\prime = \frac{E^\prime}{m_0}$. Since $E^\prime$ is the deformed total energy of the particle in the $\kappa$-deformed space-time, this leads to deformation of the boost factor due to $\kappa$-deformation of the space-time. Thus we obtain the modified boost factor as
\be 
\gamma^\prime = \frac{E^\prime}{m_0}=\frac{\sqrt{|\vec{P}|^2 + m_0^2}}{m_0}=\frac{\sqrt{m^{\prime 2} v^{\prime 2} + m_0^2}}{ m_0},
\ee 
\be 
\gamma^\prime =\frac{\sqrt{\gamma^{\prime 2}m_0^2 v^{\prime 2} + m_0^2}}{ m_0},\label{gamma}
\ee
where we have used
\be 
m^\prime =\gamma^\prime m_0.\label{4a}
\ee
From Eqn.(\ref{gamma}), we find
\be
\gamma^{\prime 2}=\gamma^{\prime 2} v^{\prime 2} +1.
\ee
In $\kappa$-Minkowski space-time the boost parameter $\beta^\prime =\frac{v^\prime}{c}$ and for $c=1$, we can write above equation as
\be 
\gamma^\prime =\frac{1}{\sqrt{1-\beta^{\prime 2}}}.\label{lf}
\ee
Note that all the above discussion is valid to $\it all~ orders$ in the deformation parameter. The deformed three momentum $\vec{P}$ in the $\kappa$-deformed space-time is given by $-i\vec{D}$ in the momentum space, where $\vec{D}$ is the Dirac derivative given in Eqn.(\ref{dr}). Thus the momentum $P_i$ is given as
\be 
P_i =-iD_i=-i\partial_i \frac{e^{-A}}{\varphi}=p_i\frac{e^{-A}}{\varphi},
\ee
where $p_i=-i\partial_i$ and $A=-ia\partial_0$. Thus we find the momentum of particle in the $\kappa$-deformed space-time as
\be 
m^\prime v^\prime =m v\frac{e^{-A}}{\varphi},\label{m}
\ee
\be 
\frac{m_0}{\sqrt{1-v^{\prime 2}}} v^\prime = \frac{m_0}{\sqrt{1-v^{2}}}v\frac{e^{-A}}{\varphi}.\label{mu}
\ee
In the above $m^\prime$ (see Eqn.(\ref{4a})) and $m$ are the relativistic mass of the particle in $\kappa$-Minkowski space-time and commutative space-time, respectively. For the realisation $\varphi =e^{-\frac{A}{2}}$, the velocity of particle in $\kappa$-Minkowski space-time, valid to all order in deformation parameter obtained from Eqn.(\ref{mu}) is 
\be 
v^\prime =\frac{ve^{-\frac{aE}{2}}}{\sqrt{1-v^2(1-e^{-aE})}}, 
\ee
where $E =p_0= \sqrt{|\vec{p}|^2 + m_0^2}$ is the total energy of the particle in the commutative space-time and $\vec{p}$ is the momentum of the corresponding particle. This shows that the boost factor in Eqn. (\ref{lf}) is modified due to $\kappa$-deformation of space-time. In limit $a\rightarrow 0$, we get back the usual boost factor in commutative space-time.

\section{Undeformed $\kappa$-Lorentz Transformation}

In this section, we derive the finite transformations in $\kappa$-Minkowski space-time. For the sake of simplicity and clarity, we consider here a frame of reference moving along the $x$-axis with respect to the another frame of reference in $\kappa$-Minkowski space-time. Using Eqn.(\ref{dg2}) for the realization $\varphi=e^{-\frac{A}{2}}$, we find the boost generators as
\begin{equation}
M_{i0}=x_i \partial_0 -x_0 \partial_i-\frac{ia}{2}x_i \partial_0 \partial_0+\frac{ia}{2}x_0 \partial_i \partial_0+\frac{ia}{2}x_i \partial_k \partial_k-\frac{ia}{2}x_k \partial_k \partial_i .\label{dg3}
\end{equation}
Now using this boost generator, we obtain the relation between prime coordinates and unprimed coordinates (assuming that the transformation is along $x$-axis) as
\be
x^\prime =  M_{10}~ x =t,\label{lt1} 
\ee
\be
t^\prime = M_{10} ~t=x. \label{lt2}
\ee
Here note that if the boost generators given in Eqn.(\ref{dg3}) have terms involving to second or higher order in deformation parameter $a$, then the above two equations will not be linear. We can re-write above two equations in the matrix form as
\be
\begin{pmatrix}
t^\prime \\
x^\prime
\end{pmatrix} 
= M_{10}\begin{pmatrix}
t \\
x
\end{pmatrix}
=\begin{pmatrix}
0 & 1\\
1 & 0
\end{pmatrix}
\begin{pmatrix}
t \\
x
\end{pmatrix}
=\sigma_x \begin{pmatrix}
t \\
x
\end{pmatrix}.
\ee
Using this we write
\be 
e^{\phi M_{10}}=e^{\phi \sigma_x}=\sum_{n=0} ^{\infty} \frac{\phi^n}{n!} (\sigma_x)^n,
\ee
Note that the effect of $\kappa$-deformation enters the above equation through the boost parameter $\phi$ only. Using this equation, we find
\be 
e^{\phi M_{10}}\begin{pmatrix}
t \\
x
\end{pmatrix}
=\left[\sum_{n=even} ^{\infty} \frac{\phi^n}{n!}I + \sum_{n=odd} ^{\infty} \frac{\phi^n}{n!} \sigma_x \right]
\begin{pmatrix}
t \\
x
\end{pmatrix},\nonumber
\ee
\be 
~~~~~~~=\left[\cosh\phi ~I + \sinh\phi ~\sigma_x \right]
\begin{pmatrix}
t \\
x
\end{pmatrix}.
\ee
Thus, we get
\be 
\begin{pmatrix}
t^\prime \\
x^\prime
\end{pmatrix}
=e^{\phi M_{10}}\begin{pmatrix}
t \\
x
\end{pmatrix}
=\begin{pmatrix}
\cosh\phi & \sinh\phi \\
\sinh\phi & \cosh\phi
\end{pmatrix}
\begin{pmatrix}
t \\
x
\end{pmatrix}
\ee
We can write above transformation in four dimensions as
\be 
\begin{pmatrix}
t^\prime \\
x^\prime \\
y^\prime \\
z^\prime
\end{pmatrix}
=\begin{pmatrix}
\cosh\phi & \sinh\phi & 0 & 0 \\
\sinh\phi & \cosh\phi & 0 & 0 \\
0 & 0 & 1 & 0 \\
0 & 0 & 0 & 1 
\end{pmatrix}
\begin{pmatrix}
t \\
x \\
y \\
z
\end{pmatrix}
\equiv B\begin{pmatrix}
t \\
x \\
y \\
z
\end{pmatrix},\label{fn}
\ee
where $B$ stands for the boost matrix. Note that the above transformation is the finite transformation along the $x$-axis. Now we parametrise above equation by using
\be 
\cosh\phi =\gamma^\prime , ~~\sinh\phi = \gamma^\prime\beta^\prime ,\label{gm}
\ee
where $\gamma^\prime$ is given in Eqn(\ref{lf}). Note that deformed parameter $\gamma^\prime$ and $\beta^\prime$ are valid to all orders in the deformation parameter. Using this we can re-write Eqn.(\ref{fn}) as
\be 
\begin{pmatrix}
t^\prime \\
x^\prime \\
y^\prime \\
z^\prime
\end{pmatrix}
=\begin{pmatrix}
\gamma^\prime & \gamma^\prime\beta^\prime & 0 & 0 \\
\gamma^\prime\beta^\prime & \gamma^\prime & 0 & 0 \\
0 & 0 & 1 & 0 \\
0 & 0 & 0 & 1 
\end{pmatrix}
\begin{pmatrix}
t \\
x \\
y \\
z
\end{pmatrix}.
\ee
From Eqn.(\ref{gm}) it is clear that $\gamma^{\prime 2} - \beta^{\prime 2}\gamma^{\prime 2} = 1$ or $\gamma^\prime =\frac{1}{\sqrt{1-\beta^{\prime 2}}}$. Here, we have taken $c=1$. Using this we can write above equations in a compact form as
\be
t^\prime = \gamma^\prime \left(t+\beta^\prime x\right),
\ee
\be
x^\prime = \gamma^\prime \left(x+\beta^\prime t\right),\label{lt3} 
\ee
\be
y^\prime = y,~~z^\prime = z. \label{lt4}
\ee
Note that the $\gamma^\prime$ appear in the above equation is given by $(1-\beta^{\prime 2})^{-\frac{1}{2}}$ and this contain the all information about $\kappa$-deformation. Thus these are the Lorentz transformation in $\kappa$-Minkowski space-time valid to all order in the deformation parameter $a$. In limit $a\rightarrow 0$, we get back the well known result in commutative space-time. Note that all the effect of $\kappa$-deformation (valid to all order in the deformation parameter $a$) is contained in $\beta^\prime$ (and $\gamma^\prime$).

\section{Dirac Equation in $\kappa$-Minkowski space-time}

In this section, using the $\kappa$-deformed Lorentz algebra we derive the Dirac equation in $\kappa$-Minkowski space-time. Since $\gamma^\prime = \cosh\phi, ~ \gamma^\prime\beta^\prime = \sinh\phi$ then the boost transformation in terms of variable $\phi$, with
\be 
\tanh\phi=\frac{ve^{-\frac{aE}{2}}}{\sqrt{1-v^2(1-e^{-aE})}}.\label{27} 
\ee 
given in Eqn.(\ref{fn}). Note that the deformation parameter $a$ is contained in the definition of $\phi$. As in commutative space-time, the boost generator along $x$-axis defined as
\be
K_x =\left. \frac{1}{i}\frac{\partial B}{\partial \phi}\right|_{\phi =0} \nonumber 
\ee
Note that the boost matrix $B$ in Eqn.(\ref{fn}) is parametrised in terms of $\phi$ and this $\phi$ which captures the effect of the $\kappa$-deformation is valid to all orders in the deformation parameter. Explicitly, we find
\be 
K_x=-i\begin{pmatrix}
0 & 1 & 0 & 0\\
1 & 0 & 0 & 0\\
0 & 0 & 0 & 0\\
0 & 0 & 0 & 0
\end{pmatrix}.
\ee
Using this procedure, we also obtain the boost matrices along the $y$ and $z$-axis as
\be
K_y = -i
\begin{pmatrix}
0 & 0 & 1 & 0\\
0 & 0 & 0 & 0\\
1 & 0 & 0 & 0\\
0 & 0 & 0 & 0
\end{pmatrix},~~
K_z = -i
\begin{pmatrix}
0 & 0 & 0 & 1\\
0 & 0 & 0 & 0\\
0 & 0 & 0 & 0\\
1 & 0 & 0 & 0
\end{pmatrix}.
\ee 
Since the rotation generators of undeformed $\kappa$-Poincare algebra is not modified in comparison with the corresponding generators of usual Poincare algebra, we get the rotation generators as in the commutative space-time. Thus we get
\be
J_x = -i
\begin{pmatrix}
0 & 0 & ~~0 & 0\\
0 & 0 & ~~0 & 0\\
0 & 0 & ~~0 & 1\\
0 & 0 & -1 & 0
\end{pmatrix},~~
J_y = -i
\begin{pmatrix}
0 & 0 & 0 & ~~0\\
0 & 0 & 0 & -1\\
0 & 0 & 0 & ~~0\\
0 & 1 & 0 & ~~0
\end{pmatrix},\nonumber
\ee 
\be
J_z = -i
\begin{pmatrix}
0 & ~~0 & 0 & 0\\
0 & ~~0 & 1 & 0\\
0 & -1 & 0 & 0\\
0 & ~~0 & 0 & 0
\end{pmatrix}.
\ee
These six generators (three for boost and three for rotation) satisfy the commutation relations
\be  
\left[K_i,~K_j\right] = -i\epsilon_{ijk}J_k~~ \label{cm1}
\ee
\be 
\left[J_i,~K_j\right] = -i\epsilon_{ijk}K_k,
\ee
\be 
\left[J_x,~K_x\right] =0,~~~~~~~~~~
\ee
\be 
\left[J_i,~J_j\right] = i\epsilon_{ijk}J_k,~~~\label{cm2}
\ee
define the undeformed $\kappa$-Lorentz algebra. Note that these matrices are exactly same as in the commutative case. All the effect of noncommutativity is included in $\beta^\prime$ and thus in $\phi$. As in commutative case, we can re-express the above algebra in Eqns.(\ref{cm1}-\ref{cm2}) in a different basis. For this we introduce $K$ matrices in terms of Pauli matrices as
\be 
K = \pm i \frac{\sigma}{2}. \label{pauli}
\ee
We define the generators of Lorentz algebra as linear combination of $J$ and $K$ as
\be 
A=\frac{1}{2}\left(J+iK\right),
\ee
\be
B=\frac{1}{2}\left(J-iK\right).
\ee
Using Eqns.(\ref{cm1}-\ref{cm2}), we find that these generators satisfy the algebra
\be  
\left[A_i,~A_j\right] = i\epsilon_{ijk}A_k,~~~~~~~~~~~~~~~~~~
\ee
\be 
\left[B_i,~B_j\right] = i\epsilon_{ijk}B_k,~~~~~~~~~~~~~~~~~~
\ee
\be 
\left[A_i,~B_j\right] =0,~~(i,~j=x,~y,~z).
\ee
Thus we see that these two sets of generators $A$ and $B$ generate $su(2)$ algebra and these two generators commute with each other showing that the $\kappa$-Lorentz algebra can be written as $su(2)\otimes su(2)$. Now we denote the angular momentum corresponding to the $su(2)$ generated by $A$ and $B$ as $j_A$ and $j_B$, respectively. For a special cases where $B=0$, i.e. $(j_A,~0)$ and $J^{(j)}=iK^{j}$, and for $A=0$, i.e. $(0,~j_B)$ and $J^{(j)}=-iK^{j}$. These two leads to the definition of two types of spinors. For $\left(\frac{1}{2},~0\right)$ representation, we have  
\be 
J^{(1/2)} = \sigma/2, ~~~K^{(1/2)} = -i\sigma/2,\nonumber
\ee
and we denote the corresponding spinor as $\xi$. If $(\theta,~\phi)$ are the rotation and pure Lorentz transformation parameters, respectively, then $\xi$ transform as
\be 
\xi \longrightarrow \exp \left[i\frac{\sigma}{2}\cdot(\theta -i\phi)\right]\xi.\label{45}
\ee
Note that the boost parameter $\phi$ contains the effect of deformation, valid to all orders in the deformation parameter $a$. And for $\left(0,~\frac{1}{2}\right)$ representation, we have  
\be 
J^{(1/2)} = \sigma/2, ~~~ K^{(1/2)} =i\sigma/2, \nonumber
\ee
and the corresponding spinor-$\eta$ transforms as
\be 
\eta \longrightarrow \exp \left[i\frac{\sigma}{2}\cdot(\theta +i\phi)\right]\eta.\label{46}
\ee
Under the $\kappa$-Lorentz transformation these spinors transform as
\be 
\Psi=
\begin{pmatrix}
\xi\\
\eta
\end{pmatrix}
\longrightarrow
\begin{pmatrix}
\exp \left[i\frac{\sigma}{2}\cdot(\theta -i\phi)\right] & o\\
0 & \exp \left[i\frac{\sigma}{2}\cdot(\theta +i\phi)\right]
\end{pmatrix}
\begin{pmatrix}
\xi\\
\eta
\end{pmatrix}.
\ee
From this, we see that under the pure $\kappa$-Lorentz transformation (when the rotation parameter $\theta =0$) the spinors $\xi$ and $\eta$ transform as
\be 
\xi \longrightarrow \phi_R ,~~\eta\longrightarrow \phi_L,
\ee
where $\phi_R$ is the right spinor and $\phi_L$ is the left spinor. Thus Eqn.(\ref{45}) can be written as
\be 
\phi_R \longrightarrow e^{\frac{1}{2}\sigma \cdot \phi}\phi_R = \left[\cosh\frac{\phi}{2}+ \sigma \cdot n \sinh\frac{\phi}{2} \right]\phi_R,\label{48}
\ee 
where $n$ is the unit vector in the direction of Lorentz boost. Here note that the boost parameter $\phi$ is now given by Eqn.(\ref{27}) and this contains all the effects of $\kappa$-deformation. Suppose initially spinor $\phi_R(0)$ describe the particle at rest and $\phi_R(\vec{P})$ describe the particle with momentum $\vec{P}$. From Eqn.(\ref{gm}), we have $\cosh\phi=\gamma^\prime$ and we find $\cosh\frac{\phi}{2}=\left(\frac{\cosh\phi +1}{2}\right)^\frac{1}{2}=\left(\frac{\gamma^\prime +1}{2}\right)^{\frac{1}{2}}$ and $\sinh\frac{\phi}{2}=\left(\frac{\cosh\phi -1}{2}\right)^\frac{1}{2}=\left(\frac{\gamma^\prime -1}{2}\right)^{\frac{1}{2}}$. Using this in Eqn.(\ref{48}), we get
\be 
\phi_R(\vec{P})=\left[\left(\frac{\gamma^\prime +1}{2}\right)^{\frac{1}{2}}+\sigma \cdot \hat{P}\left(\frac{\gamma^\prime -1}{2}\right)^{\frac{1}{2}}\right]\phi_R(0).\label{51}
\ee
Here $\hat{P}$ is the momentum unit vector along the direction of undeformed $\kappa$-Lorentz boost. Note that the effect of $\kappa$-deformation is encoded in $\gamma^\prime$ as well as in $\vec{P}$, appearing in above equation. Now using $\gamma^\prime = \frac{E^\prime}{m_0}$ and $\hat{P}=\frac{\vec{P}}{|\vec{P}|}$ in Eqn.(\ref{51}), we obtain
\be 
\phi_R(\vec{P})=\frac{E^\prime +m_0+\sigma \cdot \vec{P}}{\sqrt{2m_0(E^\prime +m_0)}}\phi_R(0).\label{52}
\ee
Similarly, we find
\be 
\phi_L(\vec{P})=\frac{E^\prime +m_0-\sigma \cdot \vec{P}}{\sqrt{2m_0(E^\prime +m_0)}}\phi_L(0).\label{53}
\ee
In the above two equations, we have replaced $\gamma^\prime$ in terms of $E^\prime$ and $m_0$. Thus the effect of $\kappa$-deformation is now contained in $E^\prime$ and $\vec{P}$ appearing in the above two equations. Since from the Eqns.(\ref{52},\ref{53}), it is clear that for particles at rest, $\phi_R(0) = \phi_L(0)$ and this allows us to eliminate the zero momentum spinors. Thus from Eqn.(\ref{52}) and Eqn.(\ref{53}), we find (see appendix for details)
\be
\phi_R(\vec{P})=\frac{E^\prime +\sigma \cdot \vec{P}}{m_0}\phi_L(\vec{P}),\label{sigma}
\ee 
and
\be
\phi_L(\vec{P})=\frac{E^\prime -\sigma \cdot \vec{P}}{m_0}\phi_R(\vec{P}).
\ee
We re-write the above two equations as
\be 
-m_0\phi_R(\vec{P})+(P_0 +\sigma \cdot \vec{P})\phi_L(\vec{P})=0,\nonumber
\ee
\be 
(P_0 -\sigma \cdot \vec{P})\phi_R(\vec{P})-m_0\phi_L(\vec{P})=0,
\ee
where $P_0 =E^\prime$. Which in the matrix form can be written as
\be 
\begin{pmatrix}
-m_0 & P_0 +\sigma . \vec{P}\\
P_0 -\sigma . \vec{P} & -m_0
\end{pmatrix}
\begin{pmatrix}
\phi_R(\vec{P})\\
\phi_L(\vec{P})
\end{pmatrix}=0.\label{54}
\ee
Defining the 4-spinor and $4\times 4$ matrices as
\be 
\Psi=
\begin{pmatrix}
\phi_R(\vec{P})\\
\phi_L(\vec{P})
\end{pmatrix},~~~
\gamma^0 =
\begin{pmatrix}
0 & 1\\
1 & 0
\end{pmatrix},~~~
\gamma^i =
\begin{pmatrix}
0 & -\sigma^i\\
\sigma^i & 0
\end{pmatrix},
\ee
Eqn.(\ref{54}) can be re-written as
\be 
\left(\gamma^0 P_0 + \gamma^i P_i -m_0\right)\Psi=0.\label{57}
\ee
Here, $P_0 = -iD_0$ and $P_i = -iD_i$. The explicit form of the Dirac derivatives  $D_0$ and $D_i$ given in Eqn.(\ref{dr}). Using this, we find the $\kappa$-deformed Dirac equation in position space as
\begin{equation}
\left(i\gamma^0 D_0 + i\gamma^i D_i + m_0\right)\Psi=0.\label{58}
\end{equation}
The $\gamma^\mu$ matrices appearing in the above equation are independent of the $\kappa$-deformation. They are ${\it exactly}$ same as in the commutative space-time. All effects of the $\kappa$-deformation of the space-time appear only through the Dirac derivatives-$D_\mu$ and all the $a$ dependent corrections are included in the above equation i.e., this deformed equation is valid to all orders in the deformation parameter.

The square of this $\kappa$-deformed Dirac equation gives the $\kappa$-deformed Klein-Gordon equation, which is invariant under the undeformed $\kappa$-Poincare algebra given in Eqns.(\ref{dirac4},\ref{dirac5})\cite{21}. The dispersion relation of corresponding  $\kappa$-deformed Klein-Gordon equation in momentum space is given in Eqn.(\ref{dis}). In limit $a\rightarrow 0$, we get back the well known Dirac equation in commutative space-time. 

\section{Conclusion}

In this paper, we have derived the $\kappa$-deformed Dirac equation valid to all orders in the deformation parameter. This was done by defining the left and right $su(2)$ spinors and studying their transformations. We have seen that the entire effects of $\kappa$-deformation is contained in the boost factor, characterising the undeformed $\kappa$-Lorentz boosts. After writing down the finite transformations of space-time coordinates under undeformed $\kappa$-Lorentz transformation, we obtain the corresponding matrix realisations of the generators. These matrices are independent of the $\kappa$-deformation. After this, by a change of basis, the undeformed $\kappa$-Lorentz algebra is written as the direct product of two $su(2)$ algebras. We then introduced spinors transforming under these $su(2)$ algebras. The effect of $\kappa$-deformation come into picture through these transformations as the boost parameter and the three momentum $\vec{P}$ characterising the transformations are deformed ones. Note that these two, namely, the boost parameter and the deformed three momentum used here all valid to all orders in the deformation parameter. The transformation between spinors under different $su(2)$ (appearing in the direct product) allowed as to write down the $\kappa$-deformed Dirac equation. We discussed that the Dirac equation obtained here is exact and include effects of deformation to all orders in the deformation parameter $a$. We also emphasis that the matrix realisation of the generators of $\kappa$-deformed Lorentz transformations (see section 4) are independent of the deformation parameter. Similarly, the $\gamma$ matrices appearing in the $\kappa$-deformed Dirac equation in Eqn.(\ref{57}) and in Eqn.(\ref{58}) are also independent of the deformation parameter. Thus $\sigma_{\mu\nu}=[\gamma_\mu,~\gamma_\nu]_+$ which gives the spinor representation will also be independent of the deformation parameter and the entire effect of the $\kappa$-deformation will be encoded in the parameter of the (boost) transformation.   
\begin{flushleft}
\begin{large}
\textbf{Acknowledgments}
\end{large}
\end{flushleft}
Author would like to thank E. Harikumar for suggestions and discussions. He also thank to UGC(India) for financial support through Rajiv Gandhi National Fellowship(F1-17.1/2011-12/RGNF-SC-UTT-4237).

\end{document}